 \documentclass[preprint2]{aastex}
 \usepackage{natbib}
 \usepackage{amsmath}
\usepackage{graphicx}
\usepackage{epsfig,psfrag,epic,eepic}
\usepackage{textcomp}
\usepackage{times}

\slugcomment{}

\shorttitle{High Contrast Imaging with Spitzer}
\shortauthors{Durkan et al.}

\begin{document}


\title{High Contrast Imaging with Spitzer : Constraining the Frequency of Giant Planets out to 1000 AU separations.}


\author{Stephen Durkan\altaffilmark{1}, 
Markus Janson\altaffilmark{2}, 
Joseph C. Carson\altaffilmark{3}
}

\altaffiltext{1}{Astrophysics Research Centre, School of Mathematics \& Physics, Queen's
University, Belfast BT7 1NN, UK; \texttt{sdurkan01@qub.ac.uk}}
\altaffiltext{2}{Department of Astronomy, Stockholm University, 106 91 Stockholm, Sweden}
\altaffiltext{3}{Department of Physics and Astronomy, College of Charleston, Charleston, SC 29424 , USA}


\begin{abstract}

 We report results of a re-analysis of archival Spitzer IRAC direct imaging surveys encompassing a variety of nearby stars. Our sample is generated
 from the combined observations of 73 young stars (median age, distance, spectral type = 85 Myr, 23.3 pc, G5) and 48 known 
 exoplanet host stars with unconstrained ages (median distance, spectral type = 22.6 pc, G5).
 While the small size of Spitzer provides a lower resolution than 8m-class AO-assisted ground based telescopes, which have been used for constraining the frequency of 0.5 - 13 $M_J$ planets at separations of $10 - 10^2$ AU, its exquisite infrared sensitivity provides the ability to place unmatched constraints on the planetary populations at wider separations.
 Here we apply sophisticated high-contrast techniques to our sample in order to remove the stellar PSF and open up sensitivity to planetary
  mass companions down to 5\arcsec\ separations. This enables sensitivity to 0.5 - 13 $M_J$ planets at physical separations on 
  the order of $10^2 - 10^3$ AU , allowing us to probe a parameter space which has not previously been 
 systematically explored to any similar degree of sensitivity. Based on a colour and proper motion analysis we do not record any planetary detections. 
 Exploiting this enhanced survey sensitivity, employing Monte Carlo simulations with a Bayesian approach, and assuming a mass distribution of $dn/dm \propto m^{-1.31}$, we constrain (at 95\% confidence) a population of 0.5 - 13 $M_J$ planets at separations
 of 100 - 1000 AU with an upper frequency limit of 9\%.

\end{abstract}





\section{Introduction}

The plethora of confirmed exoplanets to date is dominated by a population of planets within 5 -- 6 AU of their host star. This has been due to the success of large-scale radial-velocity \citep{Butler1996, Vogt2000, Mayor2003} and transit surveys \citep{Pollacco2006, Baglin2009, Koch2010}, which are biased to short separation planets and whose typical duration limit their ability to detect periodicities on the order of 10's of years. Such surveys account for a combined detection of $\sim1900$ planets (See http://exoplanet.eu and http://exoplanetarchive.ipac.caltech.edu/), $>90$\% of the entire exoplanet population. This sample enables statistically significant trends to be uncovered between planetary / host star properties and planet frequency, which in turn allows for planet formation and evolution theories to be stringently tested and constrained. However formation scenarios and evolutionary paths for wide giant planets with separations $>>$ 5 -- 6 AU continue to prove challenging to constrain. This is due to a lack of systematic explorations by surveys with a sufficient degree of sensitivity and statistically robust population analyses.

Direct imaging provides the most viable technique to probe for giant planets at such wide separations. Extensive work has been carried out to expand the sample of wide giant planets through detection in large-scale direct imaging surveys \citep{Masciadri2005, Biller2007, Lafreniere2007a, Kasper2007, Chauvin2010, Leconte2010, Ehrenreich2010, Carson2011, Janson2011, Vigan2012, Delorme2012, Biller2013, Chauvin2015}. Whilst these surveys typically record non-detections, several planetary mass wide companions have been found, indicating that although these types of planets may be rare \citep{Nielsen2010}, they do exist with some frequency throughout the galaxy. By subjecting these imaging surveys to statistical analysis, an upper limit on this frequency can be determined. 
Concurring frequency upper limits have been found for wide giant planets over a separation range on the order of $10^1$ - $10^2$ AU, corresponding to the parameter space at which observations are sensitive to Jupiter mass companions. These limits are shown in Table \ref{t:1}.

\begin{table*}
\caption{Planet Frequency Upper Limits}
\label{t:1}
\centering
\begin{tabular}{cccc}
\hline
Mass Range ($M_J$)& Separation Range (AU) & \shortstack{ Planet Frequency \\Fractional Upper Limit} & Study \\
\hline
0.5 - 13.0	&	50 - 250	&	0.093	&	\citep{Lafreniere2007a}       \\
0.5 - 13.0	&	25 - 100	&	0.110	&	\citep{Lafreniere2007a} 	\\
1.0 - 20.0        &     10 - 150    &      0.060        &      \citep{Biller2013}                 \\
1.0 - 13.0      &      20 - 150    &      0.100        &      \citep{Chauvin2010}             \\
\hline
\end{tabular}
\end{table*}

 Sensitivity is confined to this range due to instrumental limitations. Imaging surveys typically favour the use of adaptive optics (AO) corrected instruments on 8m class ground based telescopes. The high angular resolution afforded by such instruments provides sensitivity to planetary mass companions at small separations, down to the order of $10$ AU, inside of which the required contrast becomes unachievable as one approaches the core of the near diffraction limited point spread function (PSF). The outer sensitivity limit stems from anisoplanatism, where AO delivers poor wave front correction at increasing separation due to the different propagation paths and hence varying wave front distortion experienced by off-axis light. This substandard correction results in a decrease in image quality and therefore a reduction in sensitivity at large separations. Typical values for the isoplanatic angle, where distortion is statistically uniform allowing for good AO correction, are $10\arcsec$ - $20\arcsec$ in the near infrared (NIR) \citep{Sandler1994, Fritz2010}, and thus imaging instruments are typically restricted to this field of view (FOV).  This has severely limited sensitivity to planetary mass companions at separations beyond the order of $10^2$ AU, for typical nearby targets.

Conducting imaging surveys from space based telescopes would negate this effect, however their small aperture diameters produce large diffraction limited PSF's, severely limiting their application for direct imaging of planets. Still, \citet{Marengo2006, Marengo2009} have shown that the Spitzer space telescope is capable of sensitivity to planetary mass companions at large angular separations, within the background noise limited regime, with their studies of $\epsilon$ Eri and Fomauhaut. Recent studies of Vega, Fomalhaut and $\epsilon$ Eri by \citet{Janson2012, Janson2015} have demonstrated that sensitivity to planetary mass companions is achievable with Spitzer within the PSF noise-limited regime, with the application of sophisticated high-contrast reduction techniques. 

Therefore we implemented a sophisticated PSF subtraction technique to enhance the sensitivity of archival Spitzer imaging surveys, enabling sensitivity to planetary mass companions over a separation on the order of $10^2$ - $10^3$ AU.  This parameter space has not previously been systematically explored by surveys to a sufficient degree of sensitivity, leaving the population of giant planets at these separations poorly constrained. 

While the formation and origin of such wide orbit giants proves difficult to explain, their existence has been confirmed with the detection of several planetary mass companions e.g. 1RXS J1609-2105 b; 330 AU \citep{Lafreniere2010}, FW Tau b; 330 AU \citep{Kraus2014a}, HD106906 b; 650 AU \citep{Bailey2014}, GU Psc b; 2000 AU \citep{Naud2014}. Whilst core accretion and gravitational instability modes cannot readily explain the formation of such planets in situ \citep{Ida2004, Boss2006, Rafikov2007, Dodson2009}, and disk interactions cannot viably migrate planets out to $\sim10^3$ AU, beyond the typical confines of the disk \citep{Isella2009}, several theories have been developed in an attempt to account for planetary existence at these large orbital separations.

 One suggestion is that such a planetary system could be the result of dynamical capture of free-floating planets during dispersal of a stellar cluster. Simulations found this capable of producing planets at separations of $>50$ AU \citep{Parker2012} and $10^2$ - $10^5$ AU \citep{Perets2012}. As an alternative, planet-planet scattering, in which multiple planets gravitationally interact after disk dissipation, can result in dynamical scattering of a planet out to wide separations on the order of 100's of AU \citep{Rasio1996, Veras2004, Chatterjee2008, Juric2008}. Dynamical simulations by \citet{Veras2009} show these interactions capable of scattering giant planets out to separations of $10^2$ - $10^5$ AU. However they find the population of planets that eventually end up in unbound orbits passing through these wide separations to be larger than the population on stable orbits. The total population of wide giant planets then decreases on timescales of $\sim10$ Myr to produce a significantly depleted population at ages $>50$ Myr where the majority of planets have been ejected from the system. 

Constraining the population of giant planets at separations of $10^2$ - $10^3$ AU is essential for assessing the relevance of these theories, and allowing for stringent constraints to be placed on formation and evolution modes out to 1000's of AU.

\section{Target Sample}


Our target list is compiled from archival Spitzer data based on proximity and youth.
Such targets prove favourable to imaging searches for planetary companions as young giant planets are intrinsically bright in the near infrared due to heat retention from formation. This brightness decreases as the planet ages and the heat dissipates \citep{Baraffe2003, Burrows2003, Fortney2008}. Younger targets therefore provide imaging sensitivity to a greater range of companion masses for a given detection threshold, compared to relatively older stars, allowing for detection of lower mass planets. Stars within close proximity to Earth provide sensitivity to companions over a greater range of physical separations for a given detection threshold at a specific angular separation, compared to a relatively distant target, allowing for detection of shorter period planets for a given mass sensitivity.

This motivates our choice of stars from the archival Spitzer program 34 (P34) for enhanced imaging analysis. This program targeted 73 nearby ($<30$ pc), young stars and therefore provides an ideal basis for the sample of targets chosen here. 
Multiple age indicators including X-ray luminosity, chromospheric activity, Lithium abundance, rotation and photometric colour were originally used to select targets for P34 with ages below 120 Myr. We find these age estimates to be overly optimistic. 
Using a number of sources \citep[e.g.][]{Montes2001b, Zuckerman2004b, Torres2008, Maldonado2010, Malo2013}, we identify 55 of these targets to be members of young moving groups (YMGs) and place conservative age limits on the targets corresponding to reliable age estimates of YMGs taken from the literature; these age estimates are given in Table \ref{t:2}. We then place conservative age limits on the remaining targets combining literature age estimates encompassing several techniques \citep[e.g.][]{Barnes2007, Mamajek2008, Plavchan2009,  Tetzlaff2011, Vican2012}. The median target age is 85 Myr, although age limits range from 8 to 1050 Myr. The exception to this is HD 124498. \citet{Malo2013} identifies this target as a probable member of the $\beta$ Pictoris YMG. However \citet{Chauvin2010} use several age dating techniques to reclassify HD 124498 as an older system with an age $\geq100$ Myr. We therefore place an age of 12 Myr - 10 Gyr on HD 124498 to ensure any statistics encompassing age estimations produce conservative results. The median target distance, spectral type and H band magnitude are 23.3 pc, G5 and 5.289 respectively.

\begin{table*}
\caption{Moving Group Age Estimates}
\label{t:2}
\centering
\begin{tabular}{cccc}
\hline
YMG& Age Estimate (Myr) & Age Reference \\
\hline
Local Association (LA)	&	20 - 150 & \citep{Montes2001b, Brandt2014}      \\
$\beta$ Pictoris & 12 - 22 & \citep{Malo2013} \\
Ursa Major& 400 - 600 & \citep{King2003} \\
Castor & 100 - 300 & \citep{Barrado1998, Montes2001b} \\
IC2391 & 45 - 55 & \citep{Stauffer1997} \\
Hyades & 575 - 675 & \citep{Perryman1998} \\
Her-Lyr & 211- 303 & \citep{Eisenbeiss2013} \\
AB Dor & 70 - 120 & \citep{Malo2013} \\
Octans-Near & 30 - 100 & \citep{Zuckerman2013} \\
Argus & 30 - 50 & \citep{Malo2013} \\
TW Hydrae& 8 - 12 & \citep{Malo2013} \\
Tuc-Hor/Columba/Carina	 & 20 - 40 & \citep{Malo2013} \\
\hline
\end{tabular}
\end{table*}

We choose an additional archival Spitzer program to add to our target sample, program 48 (P48).
This program targeted 48 nearby ($<35$ pc) stars with known planetary companions, discovered via radial velocity. As such, these planets are on relatively short orbits, spanning a parameter space currently inaccessible to direct imaging techniques. These exoplanet host stars are relatively old and no confident age limits can be found for the majority of the sample throughout the literature. For consistency, and to justify that any derived statistical result represents a conservative limit, we adopt a conservative age of 1 Gyr - 10 Gyr for all P48 stars. This age range spans the breadth of poorly constrained literature ages for the majority of the sample.  The exceptions to this are HD 13507, HD 1237 and AF Hor, which are identified as members of YMGs with independent literature age estimates that support the association membership. Whilst these targets may prove unfavourable to a deep imaging study, due to their extended ages, the strength in their addition to the sample is that they provide an additional 48 references to aid in PSF reduction of the 73 P34 stars, contributing to an increase in achievable contrast and sensitivity to smaller mass companions. P48 was executed under the exact observational parameters as P34. Along with P48 encompassing a similar spectral sample of stars as P34, median spectral type and H band magnitude G5 and 4.957 respectively, this ensures P48 targets provide sufficient P34 PSF references. The median distance of P48 stars is 22.6 pc. So whilst the age range of P48 stars limit sensitivity to low mass planets, their proximity ensures sensitivity to larger mass planets over a wide range of separations. Therefore their inclusion our sample for reduction and analysis as well as providing additional references is justified. However we note the potential bias introduced to any population constraint statistically derived from a deep imaging search encompassing a sample of stars hosting short period planets. If we consider planet-planet scattering to be a relevant mechanism for giant planet production at $10^2$ - $10^3$ AU separations, the presence of a detected planet at a separation comparable to where a wide giant planet is expected to initially form, incorporates some degree of bias into our statistics that we cannot quantitatively evaluate. Therefore we do not correct for this possible bias.
The combined target properties are given in Table \ref{t:3}.




\section{Observations and Data Reduction}

The combined 121 targets were observed with the Infrared Array Camera \citep[IRAC;][]{Fazio2004} on the Spitzer Space telescope \citep{Werner2004} under archival programs 34 and 48 between 2003 and 2004, during the cryogenic phase of the Spitzer mission. Images were obtained simultaneously at 3.6, 4.5, 5.8 and 8 $\mu$m (channels 1 - 4 respectively). We choose Channel 2 as the primary channel to analyse based on evolutionary model predictions that planetary mass companions are at their peak luminosity within the 4.5 $\mu$m band (Barrafe et al. 2003). Therefore this channel offers the best sensitivity for planet detection. We also analyse IRAC channel 1 images as a means to vet potential companions.

\begin{figure*}[p]
\centering
\includegraphics[width=12cm]{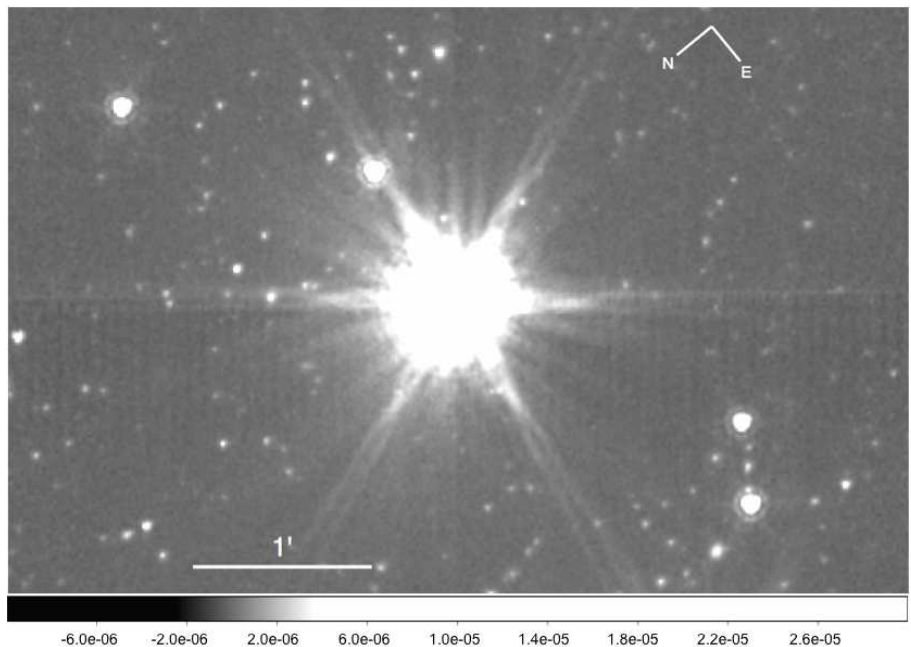}
\caption{Example of a composite PBCD / BCD 4.5 $\mu$m image of HD 217813, displaying the extent of the PSF and Spitzer spider features. The wide FOV reveals a multitude of point sources which must be vetted for planet candidacy. HD 217813 properties lie close to the median of the sample, H mag = 5.232, G5V star at 24.7pc.}
\label{f:tex}
\end{figure*}

All targets were observed with a 30s frame time allowing for an effective exposure time of 26.8s per dither position. Each observation was carried out with a five-position Gaussian dither pattern.
The Spitzer Science Center (SSC) IRAC Pipeline (version S18.25.0) performed data reduction for all observations, producing basic calibrated data (BCD) frames. This pipeline also produced corrected BCDs (CBCD) where saturated point sources have been fitted with unsaturated point sources by fitting an appropriate PSF that is matched to the unsaturated wings of the source. A final post basic calibrated data  (PBCD) frame is then created by mosaicking the relevant CBCD frames at each dither position to produce a sub-pixelated image with a $0.6\arcsec$ per pixel resolution and a $5.2\arcmin$ x $5.2\arcmin$ FOV. Such saturation corrected images are unsuitable for a close companion search as pixels towards the PSF core may have been replaced with model pixel values, ensuring that any information about potential  companion sources at these separations is lost. However we favour the use of PBCD's over BCD's due to the additional artifact correction performed on the former by the SSC IRAC pipeline. Therefore we generate a composite frame consisting of a PBCD image where saturation corrected regions, identified in the relevant mask files, have been replaced with pixel values generated from mosaicking the relevant BCD frames. These pixel values have been sub-pixelated to the equivalent $0.6\arcsec$ per pixel resolution. 
 Figure \ref{f:tex} shows an example of such a composite PBCD / BCD image.
 
\begin{figure*}[p]
\centering
\includegraphics[width=12cm]{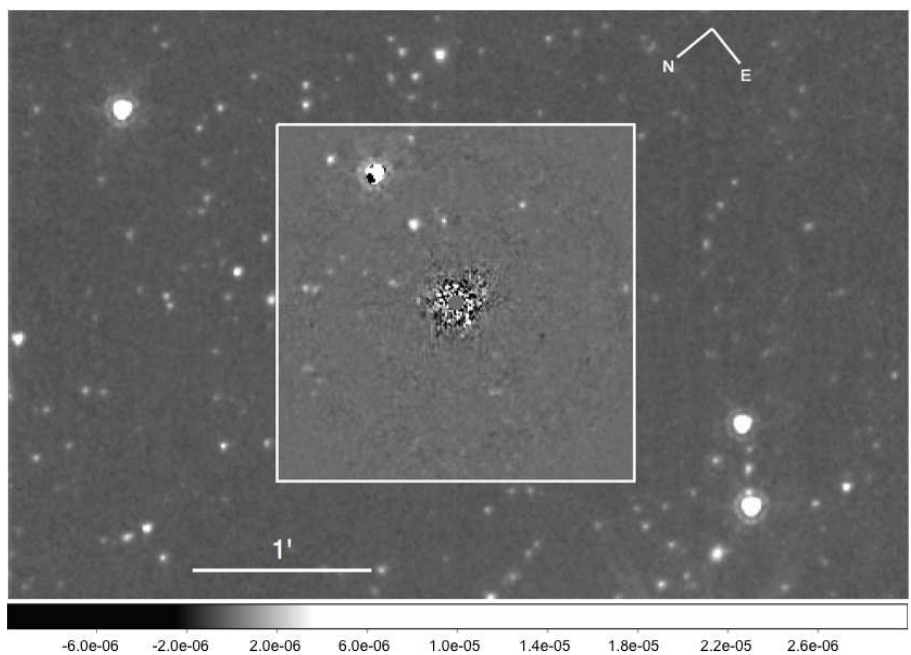}
\caption{Final reduced image of HD 217813 depicting the same FOV as Figure \ref{f:tex}. The white box highlights the $2.01\arcmin$ x $2.01\arcmin$ PCA optimization region. Regular PSF subtraction of a mean stack image has been performed outside the optimisation region.   }
\label{f:red_im}
\end{figure*}

\subsection{PSF Subtraction}
\begin{figure*}
\centering
\includegraphics[width=\textwidth]{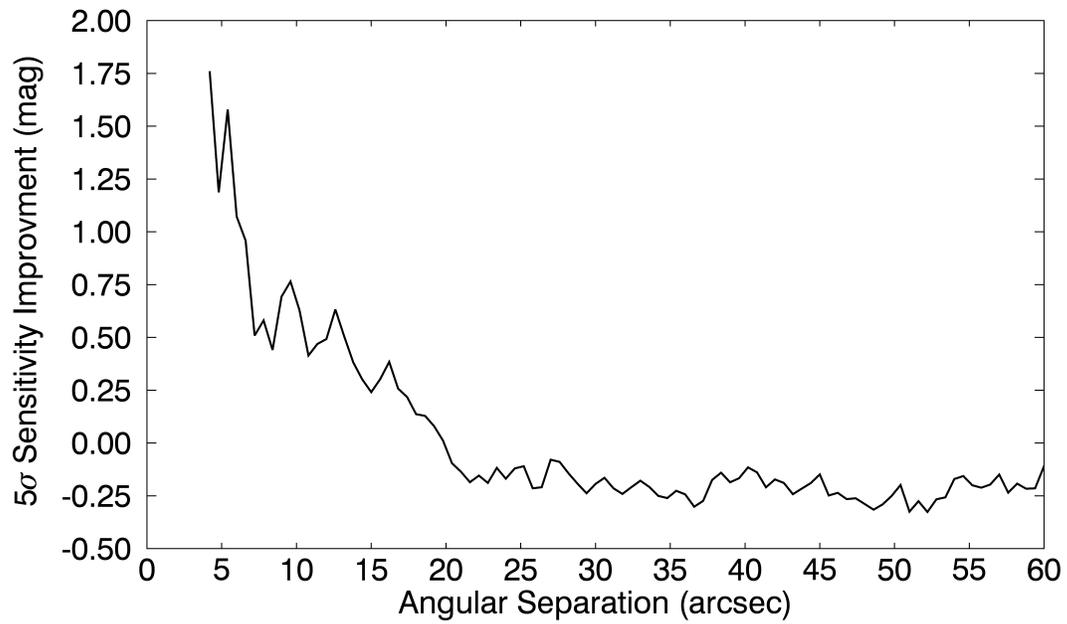}
\caption{ PCA sensitivity improvement with respect to conventional PSF subtraction, as a function of separation. The decrease in improvement around 20\arcsec\ corresponds to the transition between the PSF noise-limited regime and the background noise limited regime. The jagged nature of the curve stems from preferential reduction of inner annulus regions.  }
\label{f:improv}
\end{figure*}

We calculate the centre of each target PSF by the fitting of a 2 dimensional Gaussian, the centre of each image is then shifted to this position. Each image is then normalised by the central brightness of the stellar PSF, estimated from the saturation corrected images.  This provides an image stack of target PSF's that have similar normalised flux values. We then mask the saturated cores of each image for subsequent image reduction. Each individual image is then robustly centred with respect to the stack by subtraction of a median stack image over a range of sub-pixel offsets. The offset producing the minimum residuals is then used to align each image. The image stack constitutes the library of PSF's used to construct a reference PSF for image subtraction. This accurate alignment of target PSF's then provides better performance of any reference construction, ultimately leading to improved sensitivity in any final reduced image. 

Here we use Principal Component Analysis (PCA) to construct the optimal reference PSF to subtract from the target PSF. Variations of PCA such as PynPoint \citep{Amara2012} and KLIP \citep{Soummer2012} have the same underlying principle; a linear combination of orthogonal basis sets are used for reference construction. These basis sets represent the decomposition of the reference library into its principal components. The linear coefficients are then generated by the projection of the target onto each individual basis. PCA  performed here follows a KLIP-based analysis.

We limit the PCA optimization to a 201 x 201 pixel sub-section of each 3.6 $\mu$m and 4.5 $\mu$m image, centred on the star. This corresponds to $2.01\arcmin$ x $2.01\arcmin$ FOV. This reduced area is chosen as a favourable trade off between algorithm efficiency and sensitivity to wide separations, with $2.01\arcmin$ corresponding to separations on the order of $10^3$ AU, at the typical target distance. 
PCA is performed on concentric annuli centred on the star. The radii of the annuli are chosen such that each annulus contains 1500 pixels. Reference annuli containing astronomical sources such as background stars, stellar companions or surviving bad pixels are excluded to prevent the algorithm from subtracting any true planetary PSF , and to prevent any fake planetary signal being injected into the final image.
Since flux increases towards the PSF core, the larger flux values towards the inner region of each annulus dominate the weighting of the orthogonal basis. This may produce the lowest residuals but the outer annular regions will have experienced a substandard reduction. By performing PCA four times on each target and increasing the radius of the inner saturation mask, and therefore shifting the radii of the 1500 pixel annuli, we ensure that each section of the frame has been encompassed by an inner annular region and experienced a full quality reduction. The final images to be analysed are then composites of these four separate PCA reduced frames whose constituent parts have experienced optimal reduction. A final PCA reduced image can be seen in Figure \ref{f:red_im}.

PCA reduction is typically performed at neighbouring states of the telescope / instrument system and provides optimal PSF subtraction when carried out in tandem with angular differential imaging application, where the PSF remains quasi-static over the observation.  Here we apply PCA to a sample observed at varying states of the telescope / instrument system taken over the course of one year, e.g. varying stellar magnitude, spectral type, telescope roll angle. Therefore, our study constitutes a trial of PCA on an unfavourable data set, testing its effectiveness at the limits of its application. Similar conditions are required for optimal PSF subtraction using LOCI \citep{Lafreniere2007b} and thus we favour PCA for its speed enhancement.

\subsection{ Image Sensitivity \& Companion Identification}

We evaluate the sensitivity in the final PCA reduced images at $5\sigma$ detection limits by relating the standard deviation in concentric 1 pixel width annuli centred on the star, to the zero-point flux of Vega using the method of \citet{Marengo2009}. This generates sensitivity values in units of Vega magnitudes. As any potential companion will suffer from partial flux subtraction during the PCA reduction, we must account for the throughput of sources in the sensitivity estimation. We inject synthetic companions into each target image at position angles $0^{\circ}$ and $90^{\circ}$ so as to cover spider and non-spider regions, at five pixel intervals out to the edge of the optimisation region, and estimate throughput after PCA application.
The mean throughput is then calculated at each separation and linearly interpolated to produce throughput values for each pixel separation, consistent with sensitivity estimation in one-pixel width annuli. The mean throughput over the image stack is then used for sensitivity estimation.

We evaluate the performance of the PCA  reduction by computing sensitivity curves for both PCA optimized and conventional PSF subtracted images (i.e., subtraction of a mean stack image) and calculating the sensitivity improvement provided by PCA for each target. The median sensitivity improvement curve is then generated. This curve, displayed in Figure \ref{f:improv}, shows that PCA offers superior image sensitivity at separations less than $20\arcsec$, within the PSF noise-limited regime, and provides a median improvement of $\sim0.9$ magnitudes at separations less than $10\arcsec$. This validates the application of PCA in a study encompassing a diverse sample of stars, observed at varying states of the telescope / instrument system. The $\sim0.9$ magnitude improvement in the contrast limited regime demonstrates the relevance of applying and developing sophisticated high-contrast techniques to archival Spitzer data, where we have enhanced sensitivity to planetary mass companions at relatively small angular separations. This allows for the possibility of subsequent planet detection within a previously elusive parameter space and enabling more stringent constraints to be placed on the wide giant population. In the background noise limited regime, separations greater than $20\arcsec$, PCA does not provide any sensitivity improvement over regular PSF subtraction. This is due to the random nature of background noise which PCA cannot reference.

\begin{figure*}
\centering
\includegraphics[width=12cm]{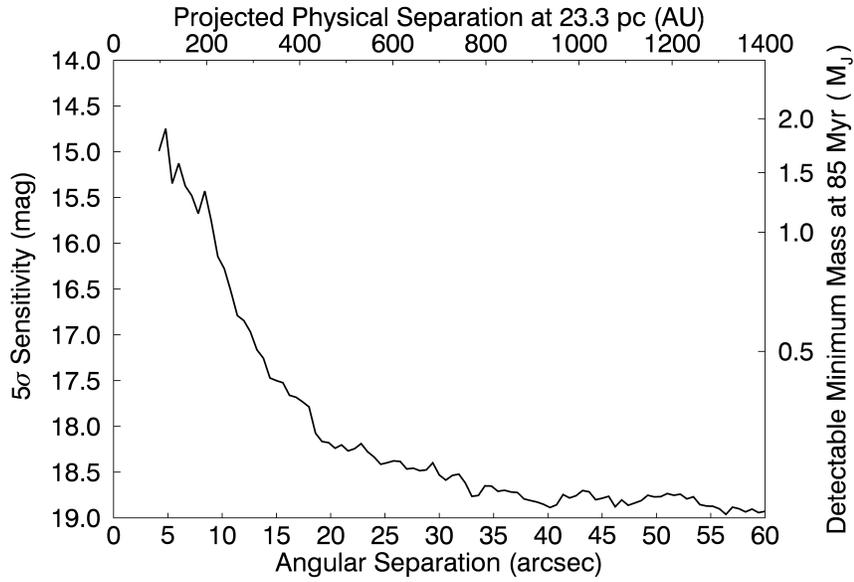}
\caption{Median survey detection limits for P34 stars. Magnitude sensitivity is given as a function of angular separation. Right axis shows corresponding minimum detectable mass at median target age of 85 Myr. Top axis displays projected physical separation at median target distance 23.3 pc.}
\label{f:p34ms}
\end{figure*}

\begin{figure*}
\centering
\includegraphics[width=12cm]{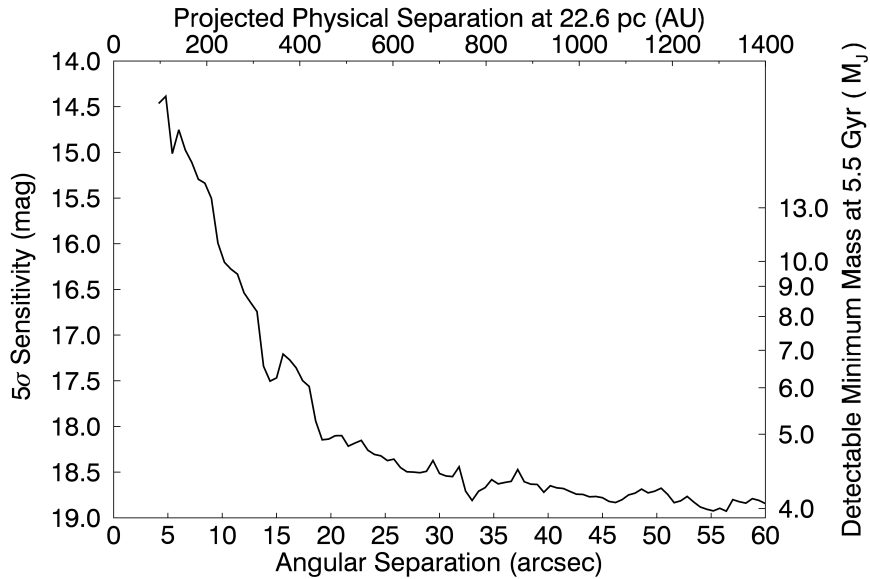}
\caption{Median survey detection limits for P48 stars. Right axis shows corresponding minimum detectable mass at median target age of 5.5 Gyr. Top axis displays projected physical separation at median target distance 22.6 pc.   }
\label{f:p48ms}
\end{figure*}

We use two initial criteria to identify potential planetary mass companions. Firstly each 4.5 $\mu$m reduced image is visually inspected to identify realistic PSF shapes, allowing real sources to be distinguished from potential surviving bad pixels. Any realistic PSF is then vetted for planetary candidacy by comparison of the 4.5 $\mu$m and 3.6 $\mu$m images. The spectral energy distribution of a non-irradiated 0.5 - 13 Jupiter mass companion, at the typical ages of our sample, is such that planetary flux at $3.6\mu$m is typically $>1$ magnitude fainter than at 4.5 $\mu$m where peak emission occurs \citep{Baraffe2003, Spiegel2012}. In contrast to a background star, which is expected to be approximately equally bright in the 3.6 $\mu$m and 4.5 $\mu$m images, a planet detection recorded around the $5\sigma$ limit at 4.5 $\mu$m is not expected to be recovered to any reasonable significance at $3.6\mu$m. This allows planetary sources to be distinguished from faint background stars without the need for a proper motion analysis. Therefore we look for source non-detection at 3.6 $\mu$m to confirm planet candidacy.

\section{Results}
\subsection{Observational Sensitivities}

Figures \ref{f:p34ms} and \ref{f:p48ms} show the 4.5 $\mu$m median sensitivity curves for the P34 and P48 stars respectively. Within $\sim20$\arcsec\ sensitivity decreases towards the PSF core where the residual PSF noise exhibits the largest variance, limiting sensitivity to lower magnitude companions. Outside $\sim20$\arcsec\ PCA provides no sensitivity improvement and we are limited by background noise which tends to be constant, thus magnitude sensitivity at these separations is roughly constant. Figures \ref{f:p34ms} and \ref{f:p48ms} also map sensitivity as a function of projected physical separation out to $\sim1400$ AU at the median distance of P34 and P48 stars, 23.3 pc and 22.6 pc respectively.
P34 and P48 stars provide comparative magnitude sensitivity limits. 

These magnitude sensitivity limits can be translated into mass sensitivities using mass-luminosity evolutionary models. One caveat is the discrepancy between hot- \citep{Chabrier2000, Baraffe2003, Burrows2003} and cold-start \citep{Marley2007, Fortney2008} models, where the later predict much fainter planets at young ages. Here we will consider hot-start models which are consistent with existing
observational data \citep{Janson2011}. However it can be noted that the models converge on the order of 10's of Myrs for low mass planetary companions ($\leq2 M_J$) and on the order of 100's of Myr for higher mass planetary companions \citep{Spiegel2012}. Thus our choice of model will not lead to significant disparity in mass sensitivity over the complete sample at the typical ages considered. COND-based models \citep{Allard2001, Baraffe2003}, applicable for companion temperatures below 1700K, which is a relevant temperature range for 0.5 - 13 $M_J$ companions at the sample ages, are used here. 

The corresponding mass sensitivities for the median P34 and P48 magnitude sensitivities are shown in Figures \ref{f:p34ms} and \ref{f:p48ms}. With P34 targets, sensitivity down to $\leq2 M_J$ companions is achieved down to  $\sim$$5\arcsec$, corresponding to $\sim100$ AU projected separation, whilst sensitivity down to 0.5 $M_J$ companions is achieved for separations $\gtrsim15\arcsec$ ($\gtrsim350$ AU). Whilst P34 and P48 stars provide comparative magnitude sensitivity limits, the extended ages of P48 stars ensure an equivalent low mass sensitivity cannot be achieved. Sensitivity to below 4 $M_J$ is typically not acquired, and sensitivity to 5 $M_J$ planets is limited to outside 20\arcsec\ ($\sim500$ AU). Taking 13 $M_J$ as the upper mass limit for planetary objects, we are typically not sensitive to planets within 10\arcsec\ ($\sim200$ AU) for these targets.

\begin{figure*}
\centering
\includegraphics[width=\textwidth,height=8cm]{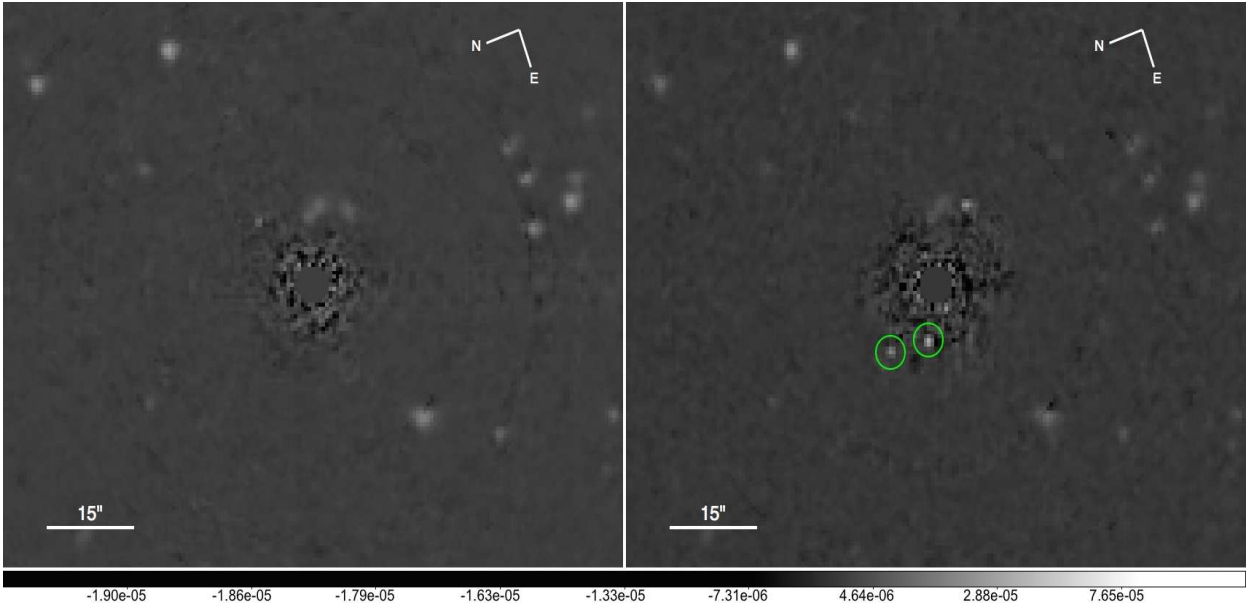}
\caption{PCA reduced 3.6 $\mu$m, left, and 4.5 $\mu$m, right, images of HD 197890. Majority of point sources revealed appear at both wavelengths and therefore are likely background stars. Sources highlighted in green circles however only appear at 4.5 $\mu$m and are identified as potential planetary candidates. After further analysis these apparent sources were revealed to be surviving outlier pixel values.}
\label{f:bomic}
\end{figure*}

\subsection{Candidate Companion Detection}

Through comparison of 3.6 $\mu$m and 4.5 $\mu$m images we initially identify 36 candidates with realistic planetary mass colour. Figure \ref{f:bomic} shows a typical candidate identification through a 4.5 $\mu$m source non-detection at 3.6 $\mu$m. These candidates are further vetted with a common proper motion analysis using a $2^{nd}$ epoch combining archival Spitzer data and observations carried out during Spitzer Cycle 11 under program 11102, repeating the original observations for several candidate host stars under the same observational parameters, with subsequent equivalent reduction. The target sample has a median total proper motion of 206 mas/yr, and over a baseline of 4 - 11 years provided by the $2^{nd}$ epoch data, we can confidently identify co-moving planetary companions in an image with a 0.6\arcsec\ /pixel scaling. Four candidates recovered in a $2^{nd}$ epoch are revealed to be non co-moving with the target star. Such sources are most likely  rare background galaxies with unusual infrared colors, such as NGC 1377 which is brighter at 4.5 $\mu$m than 3.6 $\mu$m \citep{Dale2005}. Non-detection of the remaining sources and inspection of the raw CBCD frames leads to their identification as bad pixels surviving the Spitzer Science Center IRAC Pipeline reduction. Therefore this survey records a null planet detection result.

\section{Statistical Analysis}

As in previous direct imaging surveys we exploit this null detection and the magnitude detection limits generated for each target to place constraints on the wide giant population through a coupling of Monte Carlo simulations and Bayesian analysis. 
Effectively the Bayesian analysis determines the population of wide giants, as an  upper fractional limit of stars that harbour such a companion, that is consistent with the derived planet detection probability and the null survey result. We formulate our statistical analysis in the same fashion as previous works based on \citet{Carson2006} and \citet{Lafreniere2007a}. The relevance of this work is that with improved sensitivity in the PSF-noise limited regime, due to PCA  application to archival Spitzer data, and the wide FOV providing background noise limited sensitivity, we have opened sensitivity to planetary mass companions over separations on the order of $10^2$ - $10^3$ AU. This has been done for both young, P34, and relatively old, P48, stars, as seen in Figures \ref{f:p34ms} and \ref{f:p48ms}. With this enhanced sensitivity, we can significantly constrain the population of wide giants out to 1000 AU.

\subsection{Detection Probabilities}

To derive the planet detection probability for each target we simulate 10,000 planets, using a Monte Carlo approach to randomly sample planet mass, separation, orbital projection and age. Planet mass is sampled between 0.5 - 13 $M_J$, however the lack of  constraints on the planet population over the parameter space explored here has ensured any constraints on mass distribution are correspondingly lacking. Therefore we assume a mass distribution of $dn/dm \propto m^\alpha$ where $\alpha=-1.31$, extrapolated from statistical analysis of radial velocity studies \citep{Cumming2008}. In any case \citet{Chauvin2010} find the choice of semi-major axis power law index to dominate the derived detection probabilities in comparison to any variation in $\alpha$. We choose to sample semi-major axis from a linear distribution between 100 - 1000 AU.  Studies \citep{Nielsen2008, Nielsen2010, Chauvin2010} have found semi-major axis power laws reported by \citet{Cumming2008} to be invalid at the separations considered here, motivating our choice of a linear distribution. We then sample orbital projection factor using the method of \citet{Brandeker2006} which accounts for orbital phases, orientations and an eccentricity distribution of $f(e)=2e$ \citep[predicted and observed for long period binaries;][]{Duquennoy1991}, allowing true physical separations to be translated to projected physical separations. Projected separations are then converted to angular separations using the known stellar distance. Here we do not sample over any uncertainty in stellar distance as the revised Hipparcos measurements have good precision and thus the uncertainty in stellar age  completely dominates the uncertainty in detection probabilities. Age is sampled between limits of reliable estimates from the literature, or between 1 and 10 Gyr for the majority of P48 targets with ages that are not well constrained. These ages are given in Table \ref{t:3}. Applying COND-based evolutionary models and adopting sampled planetary properties, we can translate mass into magnitude and map each planet onto $5\sigma$ magnitude sensitivity curves. The fraction of planets that lie above the detection limit provides the detection probability for each target. Mean Detection probabilities for P34 and P48 stars for mass range $[0.5, 13]$ $M_J$ and the separation range $[100, 1000]$ AU are 0.42 and 0.08 respectively. 

\begin{figure*}[p]
\centering
\includegraphics[width=14cm]{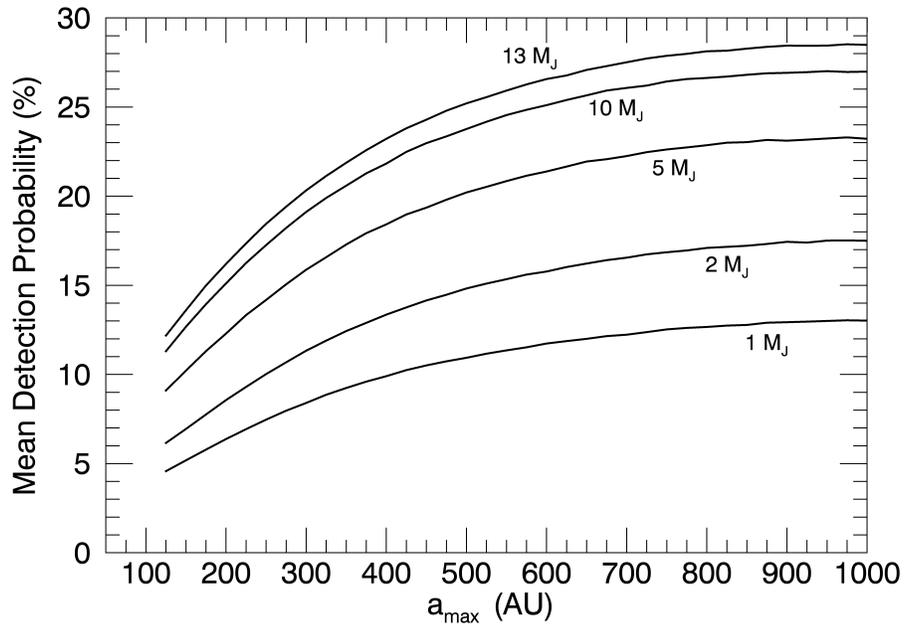}
\caption{Mean planet detection probabilities over P34 and P48 targets as a function of semi-major axis and mass. X axis denotes outer limit $a_{max}$ over which planet separation is sampled with constant $a_{min} = 100$ AU. Curves are labelled with upper mass limit $m_{max}$ over which planet mass is sampled in units of $M_J$, with constant $m_{min} = 0.5 M_J$.}
\label{f:mprobs}
\end{figure*}

We additionally choose to perform simulations over varying mass and separation ranges in order to constrain the wide giant population as a function of mass and separation. Mass range will be varied between  $[0.5, m_{max}]$ $M_J$ with $m_{max}$ ranging from 1.0 - 13.0 $M_J$ in increments of 0.5 $M_J$. Separation range will be varied between  $[a_{min}, a_{max}]$ with $a_{min}$ = 75, 100, 125 AU, and $a_{max}$ increasing in increments of 25 AU out to 1000 AU. Mean detection probabilities over the entire sample, as a function of upper mass and outer separation limit, are given in Figure \ref{f:mprobs}. Mean detection probabilities over P34 targets are typically much greater than the mean over the entire sample due to the superior mass sensitivity of P34 targets corresponding to their relatively young ages. However we find the inclusion of P48 targets provides a favourable trade off between low probability detection and increased sample size which leads to better statistical constraints on the wide giant population. 

\subsection{Binary Bias}

We identify 43 targets in our sample as binary stars in the Washington double star catalogue \citep{Mason2001} that have been vetted for true companionship through a proper motion analysis.
The inclusion of these binaries, $>1/3$ of the sample, may introduce a bias in astrophysical interpretation which we attempt to account for. We recored a null planet detection in these binary systems, which is the basis of our statistical formalism. However the binary companion will have introduced a parameter space of instability in which we would not expect a planet to orbit around its host star. To deal with this and remove the bias from our analysis we use the stability criteria of \citet{Holman1999} to determine the instability regions, for both S- and P-type systems. Many of the binary orbits and companion masses are not well documented, so for algorithm simplicity, consistency and to ensure conservative results, we take the worse-case scenario of an equal mass binary system, which introduces the greatest range of instability. We also assume the latest epoch separation in the Washington double star catalogue to be the true physical separation. This is a reasonable assumption as for a random distribution of binary eccentricities, phases and orientations the most likely true physical separation is given by the projected separation \citep{Brandeker2006}. We ensure these instability regions are counted as a non-detectable range. We note that the true frequency of wide giant planets around binary stars may be different than that around single stars. This potential frequency discrepancy is possibly due to the influence that a binary companion will have on the planet formation process. Whilst studies have found that binarity has a minimal effect on overall planet frequency \citep{Bonavita2007, Bergfors2013}, it has been suggested that binary companions with separations $\lesssim100$ AU may result in a decrease in the number of planets formed through enhanced dynamical heating of the protoplanetary disk \citep{Thalmann2014}. Whilst binary companions with separations $\lesssim100$ AU only account for 12 of the 43 binaries in our sample, an element of bias will be inherent in the statistical result derived from a combined single and binary star sample.

\subsection{Estimation of Planet Frequency}

With the determination of target detection probabilities an upper limit on the frequency of planets hosting wide giants can be derived through the Bayesian approximation; 

\begin{equation}
f_{max} \approx -ln(1-\alpha)/N\langle p_j \rangle
\end{equation}

\begin{figure*}[p]
\centering
\includegraphics[width=12cm]{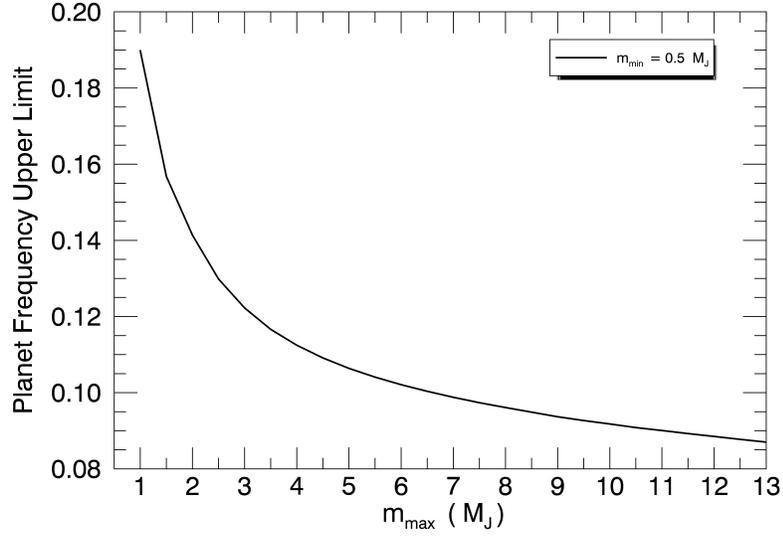}
\caption{Planet frequency upper limit (at the 95\% confidence level) as a function of mass. X axis denotes upper mass limit $m_{max}$ with constant $m_{min} = 0.5 M_J$. Separation range is constant, 100 - 1000 AU.}
\label{f:mcmass}
\end{figure*}

\begin{figure*}[p]
\centering
\includegraphics[width=12cm]{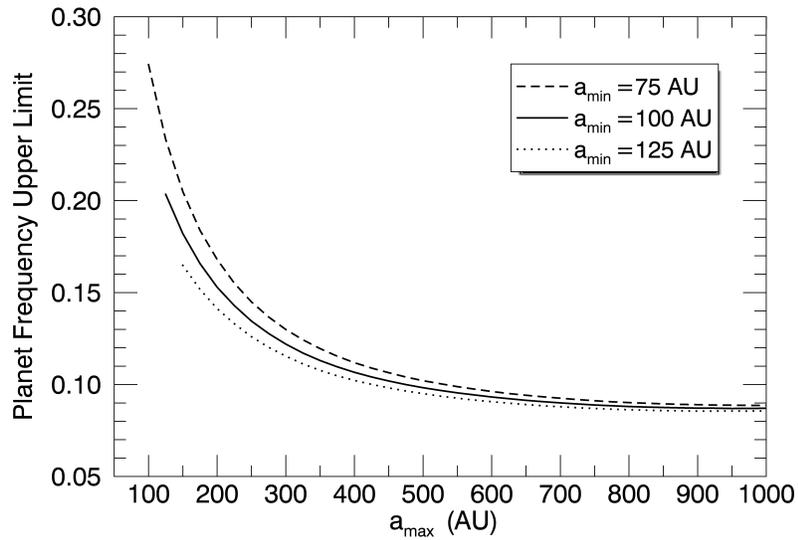}
\caption{Planet frequency upper limit (at the 95\% confidence level) as a function of separation. X axis denotes outer separation limit $a_{max}$ with constant $a_{min} = 75, 100, 125$ AU. Mass range is constant, 0.5 - 13 $M_J$.}
\label{f:mcsep}
\end{figure*}

Where $\langle p_j \rangle$ is the mean detection probability over N targets and $\alpha$ is credibility level of the derived result which we choose to be 95\%. We derive an upper limit on the frequency of planets in the mass range $[0.5, 13]$ $M_J$ and the semi-major axis range $[100, 1000]$ AU to be 9\%. As no previous survey has probed this parameter space to a similar degree of sensitivity there is no literature comparison to our result. However our low frequency findings certainly support the extension of previous survey findings of low frequency at separations on the on the order or 10 - 100 AU (see Table \ref{t:1}), out to 1000 AU separations, and is in general agreement with the theory of \citet{Veras2009}, predicting a low frequency population of $10^2 - 10^5$ AU planets at ages $>50$ Gyr. Figure \ref{f:mcmass} shows planet frequency limits as a function of mass over constant separation range 100 - 1000 AU. Figure \ref{f:mcsep} shows planet frequency as a function of separation with $a_{min}$ = 75, 100 and 125 AU over constant mass range 0.5 - 13 $M_J$. Whilst this study well constrains the population of 100 - 1000 AU planets with masses ranging from 0.5 - 13 $M_J$, the planet frequency upper limit increases towards smaller mass and separation limits. This is a deterioration of the population constraint due to bias inherent in the detection technique, where imaging sensitivity, and therefore detection probability, decreases with decreasing mass and separation. However, planet frequency tends to be a smoothly varying function with respect to mass and separation, and Figure \ref{f:mcsep} shows that at separations $\gtrsim700$ AU frequency values tend to be approximately constant, corresponding to the background noise limited regime with approximately constant sensitivity. Thus there is no reason to expect that our choice of [0.5, 13] $M_J$ and [100, 1000] AU boundaries are arbitrarily optimistic with regard to reasonably lower mass and separation limits.

\section{Conclusions}

In this paper we have presented the results of a re-analysis of two archival Spitzer imaging surveys encompassing 121 targets with varying spectral types and ages. Previously, the large PSF associated with the 0.85m Spitzer telescope diameter has severely limited its capability for directly imaging exoplanets. With the application of PCA we have removed the stellar PSF and opened up sensitivity to planetary mass companions over a broad range of separations. PCA has provided up to a magnitude sensitivity improvement at small separations with respect to conventional PSF subtraction methods, highlighting the strength of the technique, even on a relatively unfavourable data set. Using theoretical mass-luminosity evolutionary models we have shown that we are sensitive to planetary mass companions down to 0.5 $M_J$ at separations on the order of $10^2 - 10^3$ AU. This parameter space has not previously been systematically explored by imaging surveys to any comparable degree of sensitivity due to anisoplanatism and FOV limitations of ground based surveys, and PSF contrast limitations of space based surveys. Therefore through the coupling of Monte Carlo simulations and a Bayesian analysis, for the first time we have constrained the population of 0.5 - 13 $M_J$, 100 - 1000 AU planets, producing an upper frequency limit of 9\%. This is an extension of findings of low companion frequencies in numerous previous surveys at separations on the order of $10 - 10^2$ AU. Constraining this very wide giant planet population allows for previously untested formation and evolutionary theories to be adapted and constrained.

\acknowledgments
S.D. acknowledges support from the Queen's University Belfast Department for Education and Learning (DEL) university scholarship.
M.J. gratefully acknowledges funding from the Knut and Alice Wallenberg Foundation. J.C. receives support from the Research Corporation for Science Advancement (Award No. 21026) and the South Carolina Space Grant Consortium. This work is based on observations made with the Spitzer Space Telescope, which is operated by the Jet Propulsion Laboratory, California Institute of Technology under a contract with NASA. This study made use of the CDS services SIMBAD and VizieR, as well as the SAO/NASA ADS service.



\acknowledgments

\bibliography{refs}

\clearpage





\appendix


\clearpage

\begin{deluxetable}{cccccccccccc}

\tabletypesize{\scriptsize}
\rotate
\tablecaption{Target Sample Properties\label{t:3}}
\tablewidth{0pt}
\tablehead{
\colhead{HD} & \colhead{HIP} & \colhead{Other} & \colhead{RA} & \colhead{DEC} &
\colhead{H Mag} & \colhead{Spectral Type} & \colhead{Distance (pc)} &
\colhead{YMG} & \colhead{Age (Myr)} &
\colhead{YMG / Age Reference} & \colhead{Binarity} 
}
\startdata

48189	&	31711	&		&	06 38 00.366	&	-61 32 00.19	&	4.747	&	G1/G2V	&	21.3	&	AB Dor	&	70 - 120	&	Z04a, M13	&	Y/$0.6\arcsec$/$810\arcsec$	\\
	&	106231	&	LO Peg	&	21 31 01.713	&	+23 20 07.37	&	6.524	&	K8	&	24.8	&	AB Dor	&	70 - 120	&	Z04a, L06, M13	&		\\
113449	&	63742	&		&	13 03 49.655	&	-05 09 42.52	&	5.674	&	G5V	&	21.7	&	AB Dor	&	70 - 120	&	Z04a, N12, M13	&		\\
102647	&	57632	&		&	11 49 03.578	&	+14 34 19.41	&	1.925	&	A3Vvar	&	11.0	&	Argus	&	30 - 50	&	Z11, M13	&		\\
	&	23200	&	GJ 182	&	04 59 34.831	&	+01 47 00.68	&	6.450	&	M0Ve	&	25.9	&	$\beta$ Pictoris	&	12 - 22	&	T8, S12	&		\\
174429	&	92680	&		&	18 53 05.875	&	-50 10 49.88	&	6.486	&	K0Vp        	&	51.5	&	$\beta$ Pictoris	&	12 - 22	&	Z04b, T8	&		\\
196982	&	102141	&		&	20 41 51.159	&	-32 26 06.83	&	5.201	&	M4Ve	&	10.7	&	$\beta$ Pictoris	&	12 - 22	&	Z04b, T8	&	Y/$2.2\arcsec$	\\
197481	&	102409	&		&	20 45 09.531	&	-31 20 27.24	&	4.831	&	M1Ve        	&	9.9	&	$\beta$ Pictoris	&	12 - 22	&	Z04b, T8	&	Y/$4600\arcsec$	\\
181296	&	95261	&		&	19 22 51.206	&	-54 25 26.15	&	5.148	&	A0Vn        	&	48.2	&	$\beta$ Pictoris	&	12 - 22	&	Z04b, M13	&	Y/$4.2\arcsec$	\\
	&	108706	&	GJ 4247                 	&	22 01 13.125	&	+28 18 24.87	&	7.035	&	M4V	&	8.9	&	Castor	&	100 - 300	&	M01b, N12	&		\\
216803	&	113283	&		&	22 56 24.053	&	-31 33 56.04	&	3.804	&	K4V        	&	7.6	&	Castor	&	100 - 300	&	M01b, N12	&		\\
	&	114252	&	GJ 890	&	23 08 19.550	&	-15 24 35.80	&	7.301	&	M0Ve	&	22.3	&	Castor	&	100 - 300	&	M01b, N12	&		\\
13507	&	10321	&		&	02 12 55.005	&	+40 40 06.02	&	5.649	&	G5V	&	26.9	&	Castor 	&	100 - 300	&	M01b, N12	&		\\
217107	&	113421	&		&	22 58 15.541	&	-02 23 43.38	&	4.765	&	G8IV        	&	19.9	&		&	1000 - 10000	&		&		\\
206860	&	107350	&		&	21 44 31.329	&	+14 46 18.98	&	4.598	&	G0V         	&	17.9	&	Her-Lyr	&	211 - 303	&	L06, F08	&	Y/$43.2\arcsec$	\\
35296	&	25278	&		&	05 24 25.464	&	+17 23 00.72	&	4.029	&	F8V	&	14.4	&		&	84 - 316	&	B07, M08, Mk08	&	Y/$705.2\arcsec$	\\
41700	&	28764	&		&	06 04 28.440	&	-45 02 11.77	&	5.149	&	G0IV-V	&	26.6	&	Hyades	&	575 - 675	&	M01b	&		\\
173880	&	92161	&		&	18 47 01.274	&	+18 10 53.47	&	4.447	&	A5III	&	28.9	&	Hyades	&	575 - 675	&	B85	&		\\
1237	&	1292	&		&	00 16 12.678	&	-79 51 04.24	&	4.990	&	G6V         	&	17.5	&	Hyades 	&	575 - 675	&	M01b	&		\\
17051	&	12653	&		&	02 42 33.466	&	-50 48 01.06	&	4.323	&	G3IV        	&	17.2	&		&	1000 - 10000	&		&		\\
40979	&	28767	&		&	06 04 29.942	&	+44 15 37.59	&	5.509	&	F8          	&	33.1	&		&	1000 - 10000	&		&	Y/$192.4\arcsec$	\\
75732	&	43587	&		&	08 52 35.811	&	+28 19 50.95	&	4.265	&	G8V         	&	12.3	&		&	1000 - 10000	&		&	Y/$85.1\arcsec$	\\
120136	&	67275	&		&	13 47 15.743	&	+17 27 24.86	&	3.546	&	F7V         	&	15.6	&		&	1000 - 10000	&		&	Y/$1.8\arcsec$	\\
179949	&	94645	&		&	19 15 33.230	&	-24 10 45.67	&	5.101	&	F8V         	&	27.6	&		&	1000 - 10000	&		&		\\
1835	&	1803	&		&	00 22 51.788	&	-12 12 33.97	&	5.035	&	G3V	&	20.9	&	Hyades	&	575 - 675	&	M01b, M10, N12	&		\\
222143	&	116613	&		&	23 37 58.488	&	+46 11 57.96	&	5.123	&	G3/4V	&	23.3	&	Hyades	&	575 - 675	&	M01b, M10	&		\\
108799	&	60994	&		&	12 30 04.774	&	-13 23 35.46	&	4.932	&	G1/G2V      	&	24.7	&	IC2391	&	45 - 55	&	N12	&	Y/$2.1\arcsec$	\\
30495	&	22263	&		&	04 47 36.291	&	-16 56 04.04	&	4.116	&	G3V         	&	13.3	&	IC2391	&	45 - 55	&	M10	&		\\
128987	&	71743	&		&	14 40 31.106	&	-16 12 33.44	&	5.629	&	G6V	&	23.7	&	IC2391	&	45 - 55	&	M10	&		\\
19994	&	14954	&		&	03 12 46.437	&	-01 11 45.96	&	3.768	&	F8V         	&	22.6	&		&	1000 - 10000	&		&	Y/$2.1\arcsec$	\\
	&	117410	&	GJ 9839	&	23 48 25.691	&	-12 59 14.86	&	6.485	&	K5Vke	&	28.2	&	Carina	&	20 - 40	&	Z06	&		\\
27045	&	19990	&		&	04 17 15.662	&	+20 34 42.93	&	4.577	&	A3m	&	28.9	&	Octans-Near	&	30 - 100	&	Z13	&		\\
197157	&	102333	&		&	20 44 02.334	&	-51 55 15.50	&	3.692	&	A9IV	&	24.2	&	Octans-Near	&	30 - 100	&	Z13	&		\\
166	&	544	&		&	00 06 36.785	&	+29 01 17.40	&	4.629	&	K0V	&	13.7	&	Her-Lyr 	&	211 - 303	&	F04, L06	&		\\
17925	&	13402	&		&	02 52 32.128	&	-12 46 10.97	&	4.230	&	K1V	&	10.4	&	LA	&	20 - 150	&	M01a, M10	&		\\
	&	37766	&	GJ 285	&	07 44 40.174	&	+03 33 08.84	&	6.005	&	M4.5Ve	&	6.00	&	LA	&	20 - 150	&	M01b	&		\\
	&		&	EY Dra	&	18 16 16.776	&	+54 10 21.62	&	7.960	&	dM1.5e	&	30.0\tablenotemark{a}	&	LA	&	20 - 150	&	J94	&		\\
197890	&	102626	&		&	20 47 45.007	&	-36 35 40.79	&	6.930	&	K0V         	&	52.2	&	Tuc-Hor	&	20 - 40	&	K14	&		\\
10008	&	7576	&		&	01 37 35.466	&	-06 45 37.53	&	5.899	&	G5V	&	24.0	&	LA	&	20 - 150	&	M01b, M10	&	Y/$612\arcsec$	\\
77407	&	44458	&		&	09 03 27.083	&	+37 50 27.53	&	5.534	&	G0V	&	30.5	&	LA	&	20 - 150	&	M01a, M01b	&		\\
171488	&	91043	&		&	18 34 20.103	&	+18 41 24.23	&	5.896	&	G0V	&	38.0	&	LA	&	20 - 150	&	M01b	&		\\
	&		&	V383 Lac	&	22 20 07.0258	&	+49 30 11.763	&	6.577	&	K1V	&	27.5\tablenotemark{b}	&	LA	&	20 - 150	&	M01a, M01b	&		\\
130322	&	72339	&		&	14 47 32.727	&	-00 16 53.32	&	6.315	&	K0V	&	31.7	&		&	1000 - 10000	&		&		\\
217014	&	113357	&		&	22 57 27.980	&	+20 46 07.79	&	4.234	&	G5V         	&	15.6	&		&	1000 - 10000	&		&		\\
92945	&	52462	&		&	10 43 28.272	&	-29 03 51.43	&	5.770	&	K1V         	&	21.4	&	LA	&	20 - 150	&	M01b, L06	&		\\
129333	&	71631	&		&	14 39 00.210	&	+64 17 29.95	&	6.012	&	G1.5V	&	34.1	&	LA	&	20 - 150	&	M01a, M01b	&	Y/$0.8\arcsec$	\\
181327	&	95270	&		&	19 22 58.943	&	-54 32 16.97	&	5.980	&	F5/F6V      	&	51.8	&	 $\beta$ Pictoris	&	12 - 22	&	Z01, Z04b	&		\\
82443	&	46843	&		&	09 32 43.759	&	+26 59 18.70	&	5.242	&	G9V	&	17.8	&	Columba 	&	20 - 40	&	B14	&	Y/$64.7\arcsec$	\\
116956	&	65515	&		&	13 25 45.533	&	+56 58 13.78	&	5.481	&	G9V	&	21.6	&	 Her-Lyr	&	211 - 303	&	F04, M10	&		\\
177724	&	93747	&		&	19 05 24.608	&	+13 51 48.52	&	3.048	&	A0Vn	&	25.5	&	TW Hydrae	&	8 - 12	&	N12	&	Y/$1494.63\arcsec$	\\
29697	&	21818	&		&	04 41 18.856	&	+20 54 05.45	&	5.310	&	K3V	&	13.2	&	Ursa Major	&	400 - 600	&	M01a, M01b	&		\\
7590	&	5944	&		&	01 16 29.253	&	+42 56 21.90	&	5.258	&	G0          	&	23.2	&	Ursa Major	&	400 - 600	&	F04, M10	&		\\
217813	&	113829	&		&	23 03 04.977	&	+20 55 06.87	&	5.232	&	G5V	&	24.7	&	Ursa Major	&	400 - 600	&	M01b, K03	&		\\
128311	&	71395	&		&	14 36 00.560	&	+09 44 47.46	&	5.303	&	K0          	&	16.5	&		&	1000 - 10000	&		&		\\
147513	&	80337	&		&	16 24 01.289	&	-39 11 34.71	&	4.025	&	G3/G5V      	&	12.8	&		&	1000 - 10000	&		&	Y/$345\arcsec$	\\
150706	&	80902	&		&	16 31 17.585	&	+79 47 23.20	&	5.639	&	G0          	&	28.2	&		&	1000 - 10000	&		&		\\
175742	&	92919	&		&	18 55 53.225	&	+23 33 23.93	&	5.762	&	K0V	&	21.4	&	Ursa Major	&	400 - 600	&	K03	&		\\
76644	&	44127	&		&	08 59 12.454	&	+48 02 30.57	&	2.763	&	A7Vn	&	14.5	&		&	450 - 1050	&	V12	&	Y/$0.7\arcsec$/$2.4\arcsec$	\\
82558	&	46816	&		&	09 32 25.568	&	-11 11 04.70	&	5.596	&	K0V       	&	18.6	&		&	50 - 75	&	F86, T11	&		\\
92139	&	51986	&		&	10 37 18.140	&	-48 13 32.23	&	3.170	&	F4IV	&	26.8	&		&	50 - 150	&	P09	&	Y/$0.5\arcsec$	\\
112429	&	63076	&		&	12 55 28.548	&	+65 26 18.51	&	4.604	&	A5n	&	29.3	&		&	50 - 450	&	P09	&		\\
115383	&	64792	&		&	13 16 46.516	&	+09 25 26.96	&	4.107	&	G0V      	&	17.6	&		&	130 - 160	&	M10, V12	&		\\
	&		&	GJ 3789                 	&	13 31 46.617	&	+29 16 36.72	&	7.002	&	M4V	&	7.9\tablenotemark{c}	&	Carina/Columba	&	20 - 40	&	R14	&		\\
141795	&	77622	&		&	15 50 48.966	&	+04 28 39.83	&	3.440	&	A2m	&	21.6	&		&	220 - 820	&	V12	&		\\
124498	&	69562	&		&	14 14 21.357	&	-15 21 21.76	&	6.781	&	K4V         	&	30.2	&	$\beta$ Pictoris 	&	12 - 10000	&	C10, M13	&		\\
220182	&	115331	&		&	23 21 36.513	&	+44 05 52.38	&	5.574	&	K1V         	&	21.5	&		&	200-318	&	G00, B07, M08	&		\\
20630	&	15457	&		&	03 19 21.696	&	+03 22 12.72	&	3.039	&	G5Vvar      	&	9.14&		&	350-700	&	M08 , M10, V12	&		\\
130948	&	72567	&		&	14 50 15.811	&	+23 54 42.64	&	4.688	&	G2V	&	18.2	&		&	640-1001	&	D09	&	Y/$0.1\arcsec$/$2.6\arcsec$	\\
142	&	522	&		&	00 06 19.175	&	-49 04 30.68	&	4.646	&	F7V	&	25.7	&		&	1000 - 10000	&		&	Y/$4.1\arcsec$	\\
4208	&	3479	&		&	00 44 26.651	&	-26 30 56.45	&	6.243	&	G5V         	&	32.4	&		&	1000 - 10000	&		&		\\
10697	&	8159	&		&	01 44 55.825	&	+20 04 59.34	&	4.678	&	G5IV        	&	32.6	&		&	1000 - 10000	&		&		\\
3651	&	3093	&		&	00 39 21.806	&	+21 15 01.71	&	4.064	&	K0V         	&	11.0	&		&	1000 - 10000	&		&	Y/$42.9\arcsec$	\\
20367	&	15323	&		&	03 17 40.045	&	+31 07 37.36	&	5.117	&	F8V	&	26.7	&		&	1000 - 10000	&		&		\\
222404	&	116727	&		&	23 39 20.852	&	+77 37 56.19	&	1.190	&	K1III	&	14.1	&		&	1000 - 10000	&		&	Y/$0.9\arcsec$	\\
27442	&	19921	&		&	04 16 29.029	&	-59 18 07.76	&	1.814	&	K2III	&	18.2	&		&	1000 - 10000	&		&	Y/$13.1\arcsec$	\\
33636	&	24205	&		&	05 11 46.448	&	+04 24 12.73	&	5.633	&	G0          	&	28.4	&		&	1000 - 10000	&		&		\\
39091	&	26394	&		&	05 37 09.892	&	-80 28 08.84	&	4.424	&	G0V	&	18.3	&		&	1000 - 10000	&		&		\\
50554	&	33212	&		&	06 54 42.825	&	+24 14 44.02	&	5.516	&	F8V         	&	29.9	&		&	1000 - 10000	&		&		\\
75289	&	43177	&		&	08 47 40.390	&	-41 44 12.45	&	5.187	&	F9VFe+0.3	&	29.2	&		&	1000 - 10000	&		&	Y/$21.5\arcsec$	\\
82943	&	47007	&		&	09 34 50.737	&	-12 07 46.37	&	5.245	&	F9VFe+0.5	&	27.5	&		&	1000 - 10000	&		&		\\
92788	&	52409	&		&	10 42 48.528	&	-02 11 01.52	&	5.798	&	G6V	&	35.5	&		&	1000 - 10000	&		&		\\
95128	&	53721	&		&	10 59 27.973	&	+40 25 48.92	&	3.736	&	G0V         	&	14.0	&		&	1000 - 10000	&		&		\\
114386	&	64295	&		&	13 10 39.824	&	-35 03 17.21	&	6.497	&	K3V         	&	28.9	&		&	1000 - 10000	&		&		\\
114783	&	64457	&		&	13 12 43.786	&	-02 15 54.13	&	5.623	&	K1V	&	20.5	&		&	1000 - 10000	&		&		\\
192263	&	99711	&		&	20 13 59.846	&	-00 52 00.77	&	5.685	&	K2.5V	&	19.3	&		&	1000 - 10000	&		&		\\
117176	&	65721	&		&	13 28 25.809	&	+13 46 43.64	&	3.457	&	G5V         	&	18.0	&		&	1000 - 10000	&		&		\\
134987	&	74500	&		&	15 13 28.667	&	-25 18 33.65	&	5.121	&	G5V         	&	26.2	&		&	1000 - 10000	&		&		\\
141937	&	77740	&		&	15 52 17.547	&	-18 26 09.84	&	5.866	&	G2/G3V      	&	32.3	&		&	1000 - 10000	&		&		\\
143761	&	78459	&		&	16 01 02.662	&	+33 18 12.63	&	3.989	&	G2V         	&	17.2	&		&	1000 - 10000	&		&		\\
145675	&	79248	&		&	16 10 24.314	&	+43 49 03.53	&	4.803	&	K0V         	&	17.6	&		&	1000 - 10000	&		&		\\
160691	&	86796	&		&	17 44 08.701	&	-51 50 02.59	&	3.724	&	G5V         	&	15.5	&		&	1000 - 10000	&		&		\\
216435	&	113044	&		&	22 53 37.932	&	-48 35 53.83	&	4.687	&	G0V	&	32.6	&		&	1000 - 10000	&		&		\\
210277	&	109378	&		&	22 09 29.866	&	-07 32 55.15	&	4.957	&	G8V	&	21.6	&		&	1000 - 10000	&		&		\\
216437	&	113137	&		&	22 54 39.482	&	-70 04 25.35	&	4.923	&	G1VFe+0.3	&	26.8	&		&	1000 - 10000	&		&		\\
74575	&	42828	&		&	08 43 35.538	&	-33 11 10.99	&	4.227	&	B1.5III	&	269.5	&		&	13.2 - 18.4	&	T11	&		\\
59967	&	36515	&		&	07 30 42.512	&	-37 20 21.70	&	5.253	&	G3V	&	21.8	&	Castor	&	100 - 300	&	N12	&		\\
73350	&	42333	&		&	08 37 50.294	&	-06 48 24.78	&	5.318	&	G5V	&	24.0	&	Hyades	&	575 - 675	&	M10, T12	&		\\
37124	&	26381	&		&	05 37 02.486	&	+20 43 50.84	&	6.021	&	G4IV-V      	&	33.7	&		&	1000 - 10000	&		&		\\
52265	&	33719	&		&	07 00 18.036	&	-05 22 01.78	&	5.033	&	G0V	&	29.0	&		&	1000 - 10000	&		&		\\
	&		&	AF Hor	&	02 41 47.31	&	-52 59 30.7 	&	7.851	&	M2Ve	&	27.0\tablenotemark{d}	&	Tuc-Hor	&	20 - 40	&	M13, Z04b	&	Y/$22.1\arcsec$	\\
36705	&	25647	&		&	05 28 44.830	&	-65 26 54.86	&	4.845	&	K0V	&	15.2	&	AB Dor 	&	70 - 120	&	Z04a, L06, M13	&	Y/$8.9\arcsec$	\\
21845	&	16563	&		&	03 33 13.491	&	+46 15 26.53	&	6.457	&	K2	&	34.4	&	AB Dor	&	70 - 120	&	Z04a, L06, M13	&	Y/$9.6\arcsec$	\\
102077	&	57269	&		&	11 44 38.463	&	-49 25 02.75	&	6.642	&	K1V         	&	48.6	&		&	30 - 120	&	W14	&	Y/$0.2\arcsec$	\\
105963	&	59432	&		&	12 11 27.754	&	+53 25 17.45	&	5.867	&	K0V	&	30.2	&		&	242 - 302	&	B07, M08	&	Y/$13.5\arcsec$	\\
139084	&	76629	&		&	15 38 57.543	&	-57 42 27.34	&	5.994	&	K0V	&	38.5	&	$\beta$ Pictoris	&	12 - 22	&	Z04b, M13	&	Y/$10.2\arcsec$	\\
172555	&	92024	&		&	18 45 26.900	&	-64 52 16.54	&	4.251	&	A7V	&	28.6	&	$\beta$ Pictoris	&	12 - 22	&	Z04b, M13	&	Y/$71.4\arcsec$	\\
202730	&	105319	&		&	21 19 51.990	&	-53 26 57.93	&	4.224	&	A5V	&	30.3	&		&	50 - 900	&	P09, V12	&	Y/$7.3\arcsec$	\\
218738	&	114379	&		&	23 09 57.372	&	+47 57 30.13	&	5.788	&	G5Ve	&	23.7	&		&	12.7 - 48	&	P09, T11	&	Y/$15.8\arcsec$	\\
51849	&	33560	&		&	06 58 26.051	&	-12 59 30.58	&	6.368	&	K4V         	&	21.7	&		&	30 - 220	&	P09, T11	&	Y/$0.7\arcsec$	\\
141272	&	77408	&		&	15 48 09.463	&	+01 34 18.27	&	5.610	&	G8V	&	21.3	&	LA	&	20 - 150	&	M01b, M10	&	Y/$17.9\arcsec$	\\
155555	&	84586	&		&	17 17 25.505	&	-66 57 03.73	&	4.907	&	G5IV	&	31.5	&	$\beta$ Pictoris	&	12 - 22	&	Z04b, M13	&	Y/$34.04\arcsec$	\\
220140	&	115147	&		&	23 19 26.633	&	+79 00 12.67	&	5.512	&	G9V	&	19.2	&		&	16 - 50	&	Mk07, T11	&	Y/$10.8\arcsec$/$962.6\arcsec$	\\
11131	&	8486	&		&	01 49 23.356	&	-10 42 12.86	&	5.289	&	G1Vk	&	22.6	&	Ursa Major	&	400 - 600	&	M01b, M10, N12	&	Y/$192.9\arcsec$	\\
43162	&	29568	&		&	06 13 45.296	&	-23 51 42.98	&	4.863	&	G5V         	&	16.7	&	IC 2391	&	45 - 55	&	M10, N12	&	Y/$24.6\arcsec$/$164\arcsec$	\\
160934	&	86346	&		&	17 38 39.634	&	+61 14 16.03	&	6.998	&	K7Ve	&	33.1	&	AB Dor	&	70 - 120	&	Z04a, L06, M13	&	Y/$0.12\arcsec$	\\
9826	&	7513	&		&	01 36 47.842	&	+41 24 19.64	&	2.957	&	F8V	&	13.5	&		&	1000 - 10000	&		&	Y/$55.4\arcsec$	\\
13445	&	10138	&		&	02 10 25.934	&	-50 49 25.42	&	4.245	&	K0V         	&	10.8	&		&	1000 - 10000	&		&	Y/$2.4\arcsec$	\\
46375	&	31246	&		&	06 33 12.622	&	+05 27 46.53	&	6.072	&	K1IV	&	34.8	&		&	1000 - 10000	&		&	Y/$11.2\arcsec$	\\
162020	&	87330	&		&	17 50 38.355	&	-40 19 06.07	&	6.649	&	K3V         	&	29.4	&		&	1000 - 10000	&		&		\\
186427	&	96901	&		&	19 41 51.972	&	+50 31 03.08	&	4.695	&	G5V         	&	21.2	&		&	1000 - 10000	&		&	Y/$41.6\arcsec$	\\
190360	&	98767	&		&	20 03 37.406	&	+29 53 48.49	&	4.239	&	G7IV-V	&	15.9	&		&	1000 - 10000	&		&	Y/$178.2\arcsec$	\\		
\enddata

\tablecomments{ \textbf{Target right ascension (RA) and declination (DEC) values are taken from the revised Hipparcos catalogue \citep{VanL2007} at equinox=J2000, epoch=J2000 computed by Vizier.} H band magnitudes are taken from 2MASS All-Sky Catalog of Point Sources \citep{Cutri2003}. Distances taken from revised Hipparcos catalogue unless noted otherwise with indices [a,b,c,d], relevant references given below. Column 9 notes YMG membership. Age estimates for YMG targets are taken from estimates of the group age given in table 2. Targets with unconstrained 
ages are assigned an age of 1 - 10 Gyr. Column 11 notes literature sources identifying YMG membership, or the sources estimating age for several young targets with no known YMG membership. Column 12 notes Y for targets identified as part of a binary system and gives the separation of the binary companion, identification and separation values are taken from the Washington double star catalogue \citep{Mason2001}.}
\tablenotetext{a}{\citep{Plavchan2009}}
\tablenotetext{b}{\citep{Montes2001a}}
\tablenotetext{c}{\citep{Gliese1991}}
\tablenotetext{d}{\citep{Riaz2006}}
\tablerefs{(B07) \citep{Barnes2007}; (B14) \citep{Brandt2014}; (B85) \citep{Bubenicek1985}; (C10) \citep{Chauvin2010}; (D09) \citep{Dupuy2009};  (F04, F08) \citep{Fuhrmann2004, Fuhrmann2008}; (F86) \citep{Fekel1986}; (G00) \citep{Gaidos2000}; (J94)	\citep{Jeffries1994}; (K03) \citep{King2003}; (K14) \citep{Kraus2014b}; (L06) \citep{Lopez2006}; (M01a, M01b) \citep{Montes2001a, Montes2001b}; (M08) \citep{Mamajek2008}; (M10) \citep{Maldonado2010}; (M13)  \citep{Malo2013}; (Mk07, Mk08) \citep{Makarov2007, Makarov2008}; (N12) \citep{Nakajima2012}; (P09) \citep{Plavchan2009} ; (R14) \citep{Riedel2014}; (S12) \citep{Schlieder2012}; (T8) \citep{Torres2008}; (T11) \citep{Tetzlaff2011}; (T12) \citep{Tabernero2012}; (V12) \citep{Vican2012};  (W14) \citep{Wollert2014}; (Z01) \citep{Zuckerman2001}; (Z04a, Z04b) \citep{Zuckerman2004a, Zuckerman2004b}; (Z06, Z11, Z13) \citep{Zuckerman2006, Zuckerman2011, Zuckerman2013}.}

\end{deluxetable}


\end{document}